%% file: main.tex
\documentclass[conference]{IEEEtran}
\usepackage{graphicx}
\usepackage{algorithm}
\usepackage[noend]{algpseudocode}
\usepackage{tikz}

\algnewcommand\algorithmicinput{\textbf{Input:}}
\algnewcommand\INPUT{\item[\algorithmicinput]}

\ifCLASSINFOpdf
\else
\fi
\begin{document}
%
\title{Adaptive Nature-inspired Fog Architecture}

\author{\IEEEauthorblockN{Dragi Kimovski}
\IEEEauthorblockA{Distributed multimedia systems group \\
University of Klagenfurt\\
Klagenfurt, Austria\\
Email: dragi@dps.uibk.ac.at}
\and
\IEEEauthorblockN{Humaira Ijaz}
\IEEEauthorblockA{Distributed and parallel systems group \\
University of Innsbruck, Austria\\
University of Sargodha, Pakistan\\
Email: humaira.bilalrasul@uos.edu.pk}
\and
\IEEEauthorblockN{Nishant Saurabh}
\IEEEauthorblockA{Distributed and parallel systems group \\
University of Innsbruck\\
Innsbruck, Austria\\
Email: nishant@dps.uibk.ac.at} \\
\and
\hspace{6.3cm}
\IEEEauthorblockN{Radu Prodan}
\IEEEauthorblockA{
\hspace{6.3cm}
Distributed multimedia systems group \\
\hspace{6.3cm}
University of Klagenfurt\\
\hspace{6.3cm}
Klagenfurt, Austria\\
\hspace{6.3cm}
Email: radu@dps.uibk.ac.at}}

\maketitle


\begin{abstract}
During the last decade, Cloud computing has efficiently exploited the economy of scale by providing low cost computational and storage resources over the Internet, eventually leading to consolidation of computing resources into large data centers. However, the nascent of the highly decentralized Internet of Things (IoT) technologies that cannot effectively utilize the centralized Cloud infrastructures pushes computing towards resource dispersion. 
Fog computing extends the Cloud paradigm by enabling dispersion of the computational and storage resources at the edge of the network in a close proximity to where the data is generated. In its essence, Fog computing facilitates the operation of the limited compute, storage and networking resources physically located close to the edge devices. However, the shared complexity of the Fog and the influence of the recent IoT trends moving towards deploying and interconnecting extremely large sets of pervasive devices and sensors, requires exploration of adaptive Fog architectural approaches capable of adapting and scaling in response to the unpredictable load patterns of the distributed IoT applications.
In this paper we introduce a promising new nature-inspired Fog architecture, named SmartFog, capable of providing low decision making latency and adaptive resource management. By utilizing novel algorithms and techniques from the fields of multi-criteria decision making, graph theory and machine learning we model the Fog as a distributed intelligent processing system, therefore emulating the function of the human brain.

\end{abstract}


%
\IEEEpeerreviewmaketitle

\section{Introduction}
\input{sections/Introduction.tex}

\section{Terminology}
\label{sec:terminology}
\input{sections/Terminology.tex}
\section{Related work}
\label{sec:related}
\input{sections/Related.tex}
\section{Background}
\label{sec:background}
\input{sections/Background.tex}
\section{Architectural model}
\label{sec:architecture}
\input{sections/Architecture.tex}

\section{Implementation}
\label{sec:implementation}
\input{sections/Implementation.tex}
\section{Experimental evaluation}
\label{sec:evaluation}
\input{sections/Evaluation.tex}

\section{Conclusion}
\label{sec:conclusion}
\input{sections/Conclusion.tex}


\section*{Acknowledgment}
This work is being accomplished as a part of the Tiroler Cloud Project: ”Energiebewusste F{\"o}derierte Cloud f{\"u}r Anwendungen aus Industrie und Forschung”, funded by the Bridge programme under grant agreement No 848448.



%

\bibliographystyle{unsrt}
\bibliography{ref}

\end{document}

%% file: sections/Introduction.tex
%
%
%

The recent embrace of the ``Smart Anything Everywhere'' paradigm has caused a major technological tidal wave in pervasive computing, transforming the way we perceive, utilize and interact with the environment around us. Large-scale \textit{Internet of Things (IoT)} systems, such as smart cities, autonomous vehicles and intelligent health care services, are clearly the next disruptive technology encompassing various physical and virtual loosely connected devices interacting through existing communication infrastructure. IoT services are typically composed of a set of distributed components, running in different locations and connected through dynamic networks. The emergence of these technologies led to explosive growth in data generation that needs to be processed with lowest possible latency.

In the last decade, \textit{Cloud computing} has efficiently exploited the economy of scale by providing low cost computational and storage resources over the Internet, eventually leading to consolidation of computing resources into large data centers. However, the nascent of the highly decentralized IoT technologies that cannot effectively utilize the centralized Cloud infrastructures pushes computing towards resource dispersion. For example, autonomous vehicles cannot rely on the Cloud for real-time video processing or temporary data storage, as this induces unacceptably high decision making latencies. Essentially, these applications require the processing and data storage to be moved from the remote Cloud to the nearby edge of network, allowing low latency communication and processing.    

In the digital world, the connectivity provided by the Internet network can lead us to a false sense of proximity. For example, Cloud services can be perceived by the end-users as being logically in a close proximity, even though the physical distance can be stretched over different continents, resulting in higher end-to-end latency and lower available bandwidth. \textit{Fog computing} extends the Cloud paradigm by enabling dispersion of the computational and storage resources at the \textit{edge} of the network in a close proximity to where the data is generated \cite{mainEdge,2016fog,varshney2017demystifying}.
In its essence, Fog computing facilitates the operation of the limited compute, storage and networking resources physically located close to the \textit{edge devices}.
The proximity characteristics of the Fog paradigm pushes the evolution of the Cloud for the future IoT systems by enabling highly responsive services, improving the scalability to new dimensions, and providing much higher fault tolerance by masking Cloud outages \cite{orchFog}.
        

There exist various overlapping definitions of Fog computing, making it very difficult to agree on an unified architecture.  Recently, a few promising definitions of an unified Fog architecture have been proposed \cite{vaquero2014finding,fogarch}, but they are too general, omit detailed descriptions, and do not address multiple important factors, such as system scalability, interaction and communication among the Fog devices, and resource mapping. The shared complexity of the Fog and the influence of the recent IoT trends moving towards deploying and interconnecting extremely large sets of pervasive devices and sensors, aggravates this issue even more. The requirements for defining a unified Fog architecture are very high, as we deal with a complex distributed structure capable of processing a multitude of parallel heterogeneous tasks. Adding the architectural heterogeneity of the various computational devices utilized in the Fog, it is increasingly acknowledged that novel Fog architectural approaches capable of adapting and scaling in response to the external environment and distributed applications need to be explored. Therefore, we strongly believe that the highly parallel nature of the Fog can be described through an analogy with one the of most powerful and efficient processing system, the \textit{human brain}, referring to its ability to adapt its own structure and functions following the changes in the environment, a property called plasticity. 

In this paper we introduce a promising new nature-inspired Fog architecture, named SmartFog, capable of providing low decision making latency and adaptive architecture structuring. By utilizing novel algorithms and techniques from the fields of multi-criteria decision making \cite{kimovski2015parallel}, graph theory \cite{centrality} and machine learning \cite{spectral2007tutorial} we model the Fog as a distributed intelligent processing system, therefore emulating the function of the human brain. In our analogy, the Fog devices are modeled to mimic the function of the \textit{neurons}, while the \textit{synapses} are correlated with the communication channels. The Fog devices are capable of self-clustering into multiple \textit {functional areas}, optimized to support the functioning of a given IoT application and capable of restructuring in relation to the intensity and pattern of the sensory data flow. The pervasive IoT devices and sensors are represented by the sensory nervous system. For example, the temperature sensors can be related to the thermoreceptors, or the surveillance cameras to the photoreceptors. The Cloud functionality takes a twofold role in our model: (i) it will provide backbone for supporting communication between the different functional areas, therefore emulating the function of the corpus callosum, and (ii) it will enable long-term storage of the processed data sets.

The paper is organized as follows. Section~\ref{sec:related} summarizes the related research activities. Section~\ref{sec:background} provides detailed information on the related concepts and methodologies from the fields of multi-criteria decision making, graph theory, and machine learning. Section~\ref{sec:architecture} explains the architectural model and formulates the novel methodologies.  Section~\ref{sec:implementation} provides implementation details of the simulated Fog environment and the implementation of the utilized methodologies. Section~\ref{sec:evaluation} presents experimental results and Section~\ref{sec:conclusion} concludes the paper.

%% file: sections/Terminology.tex
Currently, the Fog computing paradigm is investigated as the logical evolution of the modern distributed systems (see Section \ref{sec:related}). 
Unfortunately, all these initiatives are largely fragmented, leading to multiple overlapping, interchangeable and sometimes confusing definitions of the important terms related to Fog computing.
To overcome these limitation and improve the readability of this paper, we provide the following important definitions: 
\begin{itemize}
\item \textbf{Fog environment} is a collection of interconnected smart devices and Cloud data centers, which collaboratively work to provide low-latency services closer to where the data is generated;
\item \textbf{Fog device} is an interconnected hardware usually located within the local area network, such as smart router or wireless access point that, besides serving its original purpose of supporting networking operations, can provide secondary computing and storage services to IoT applications. A Fog device can also be any dedicated physical server or small private Cloud located within the local network;
\item \textbf{Edge device} is a low-power device, such as mobile phone or wearable device, located at the edge of the network, capable of pre-processing and aggregating data from IoT sensors. 
\item \textbf{IoT application} is a modular software program that emphasizes on separating the functionalities into independent, interchangeable components, which are executed on various low-powered IoT devices for data pre-preprocessing and Cloud infrastructures for complex data operations and storage.  
\end{itemize}

%% file: sections/Related.tex
%

Recently, promising research initiatives have been started in the European research community, focused towards solving issues related to the IoT and Fog computing. One of these initiatives is the H2020 LightKone project, which aims on developing new programming models and algorithms for general-purpose computation on edge networks by incorporating a synchronization-free programming and hybrid gossip algorithms \cite{shoker2016lightkone}. Furthermore, the H2020 DITAS  project targets the development of a unified platform for data-intensive applications capable of simplifying the information logistics for Cloud and edge environments \cite{plebani2017information}.  
Furthermore, novel research activities, focused towards solving issues related to the IoT and Fog computing has been initiated in the research community. The authors in \cite{orchFog} present an intriguing early-stage concept for Fog orchestration and architecture management by implementing genetic-based heuristics, focused on gradual optimization of the IoT's workflows in relation to the QoS requirements. To achieve balance between the accuracy and time efficiency, the authors separate the computation among multiple workers and utilize a centralized master node for result aggregation and decision making. In addition to inducing a single point of failure and computational bottleneck, this approach can lead to identification of false global extremes, which influences the quality and accuracy of the results. Furthermore, in \cite{osmotic} a novel concept of Osmotic computing, which is a new paradigm for supporting an efficient execution of IoT services and applications at the network edge has been presented. This intriguing approach for Edge/Cloud computing presents interesting concepts, which could be utilized as an extension or alternative to the Fog.   

A promising orchestration model for Fog resources provisioning based on a service-oriented approach is proposed in \cite{OrchFogData}, enabling container-based resource provisioning in distributed architectures through a hybrid orchestration architecture. Unfortunately, this approach does not focus on extending the Fog architecture to consider non-functional parameters and cannot adaptively react to the changes in the workload. Similarly to the previously described concept, there is a possibility for a single point of failure if additional redundancy measures are not taken due to the centralized architecture of the orchestration module.

The work in \cite{aazam2015dynamic} proposed a novel resource management methodology for the Fog considering multiple different factors, such as user behavior, Fog devices availability, and services price. Although this approach implements intriguing concepts for resource estimation in fine-grained manner,  it does not explore methodologies adaptive architecture management. 

The work in \cite{OrchIoT} presents a conceptual framework for resource provisioning based on a centralized Cloud-Fog middleware together with a hierarchical Fog orchestration control system to manage the provisioning of the computational resources in IoT and Fog environments on a local level. However, this approach does not consider any independent communication and self-adaptation between the orchestration control nodes, therefore limiting the possibility for deployment of adaptive Fog architecture.



%% file: sections/Background.tex
%
\subsection{Multi-objective optimization}
In this work we extend and utilize essential concepts from the area of multi-criteria optimization and decision making with a main goal to enable efficient communication between the IoT devices and the Fog/Cloud systems above. In the most general sense, optimization is a process of identifying one or multiple solutions, which correspond to the extreme values of two or more objective functions within given constraints set. In the cases in which the optimization task utilizes only a single objective function it results in a single optimal solution. Moreover, the optimization can also consider multiple conflicting objectives simultaneously. In those circumstances, the process will usually result in a set of optimal trade-off solutions, so-called Pareto front. The task of finding the optimal set of Pareto solutions in the form of Pareto front is known in the literature as a multi-objective optimization~\cite{branke2008multiobjective}~\cite{kimovski2016multi}. The Pareto front is an essential tool for decision support and preference discovery, whose shape provides new insights and allows scientists to explore the space of non-dominated solutions, possibly revealing regions of interest that are impossible to see otherwise.

In general, the multi-objective optimization problem involves two or more objective functions which have to be either minimized or maximized. The problem of optimization can be formulated as:
\begin{equation}
min/max (f_1(Y), f_2(Y),\dots,f_n(Y))
\end{equation}
where $n \geq 2$ is the number of objectives functions $f$ that we want to minimize or maximize, while $Y=(y_1,y_2,\dots,y_k)$ is a region enclosing the set of feasible decision vectors.

\subsection{Betweenness centrality}
In the research field of graph theory, the  \textit{betweenness centrality} is used as a measure of node centrality within a graph, in relation to the number of shortest paths passing through the node (vertex) \cite{barthelemy2004betweenness}. For each pair of vertices in a connected unweighted graph, there exists at least one shortest path, such that the number of edges that the route passes through can be minimized. Respectively, for weighted graphs the betweenness centrality utilizes the the sum of the edges' weights to find the shortest paths. The betweenness centrality for each vertex is the number of the shortest paths that pass through the vertex and can be represented through the following equation:
\begin{equation}
g(n)= \sum_{s \neq n \neq d} \frac{\sigma_{sd}(n)}{\sigma_{sd}} 
\end{equation}
where $\sigma_{sd}$ is the total number of shortest path routes from vertex $s$ to vertex $d$, while $\sigma_{sd}(n)$ represents the number of shortest paths from vertex $s$ to vertex $d$ passing through vertex $n$.

High value of the betweenness centrality usually implies that a given vertex can reach others with lowest possible latency. Moreover, high betweenness centrality can also indicate that a given node lies on the path of many shortest routes. Consequently, if one removes a node with large centrality it can lengthen the paths between many other pairs of nodes. This concept can be utilized in Fog computing in order to find the most central Fog devices, therefore enabling more efficient communication between the IoT devices and the Fog.   

\subsection{Spectral clustering}
In the Big Data era the vital tool for dealing with large data-sets is the concept of classification or grouping of data objects into a set of categories or clusters. The classification of the objects is conducted based on the similarity or dissimilarity of multiple features that describe them. Essentially, the classification methods can be divided into two categories, namely supervised and unsupervised \cite{cluster}. In supervised classification, the features' mapping from a set of input data vectors is classified to a finite set of discrete labeled classes and it is modeled in terms of some mathematical function. On the other hand, in unsupervised classification, called clustering, no labeled data-sets are available. The aim of the clustering is to separate a finite unlabeled data-sets into a finite and discrete set of clusters. 
For the purpose of efficient clustering of the Fog devices into functional groups we utilize unsupervised clustering technique called Spectral clustering \cite{spectral2007tutorial}. Spectral clustering is a technique that make use of the eigenvalues \cite{edelman1988eigenvalues} of the similarity matrix $S_{ij}=s(x_i,x_j)$ of the data points $x_i,x_j$ to perform dimensionality reduction before clustering the data in relation to a reduced number of dimensions. The commonly used similarity measures in Spectral clustering are based on the Euclidean distance and the Gaussian kernel. After performing dimensionality reduction, Spectral clustering relies on more simple clustering techniques, such as k-means \cite{kmeans}, to group the similar data points in distinctive similarity based sub-graphs. Spectral clustering is suitable for finding resemblance between different data points by utilizing the concept of graph similarity, making it easily applicable for sub-diving of network graphs, such as the Fog environment.   

%% file: sections/Architecture.tex

\begin{figure*}
        \centering
                \includegraphics[width = 1\textwidth ]{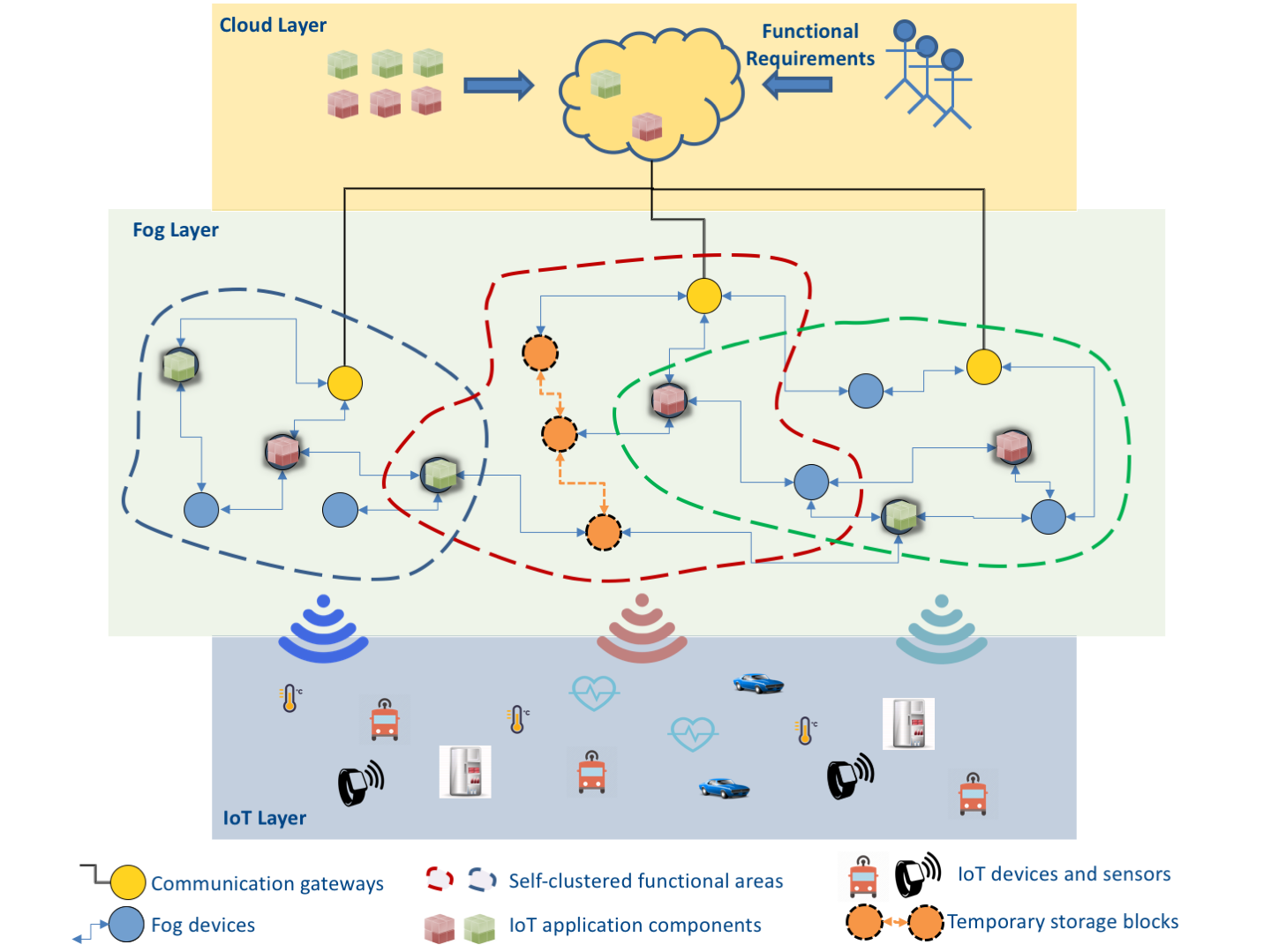}   
        \caption{SmartFog architecture}
        \label{fig:system}
\end{figure*}

The SmartFog architecture has been envisioned as a self adapting system that, similarly to the human brain, can react to environmental changes. The SmartFog architecture aims on providing a robust computational backbone to the IoT platform underneath, while efficiently utilizing the Cloud services above. The SmartFog architecture, depicted in Figure \ref{fig:system}, is composed of three distinctive layers: (1) \textit{Cloud layer}, (2) \textit{Fog layer}, and (3) \textit{IoT layer}. SmartFog loosely integrates both Cloud and IoT layers associated to the Fog, thereby enabling independent evolution and high degree of interaction.    

To provide a familiar environment to the IoT applications developers, SmartFog utilizes the Cloud as the main entry point for application deployment and execution. The application components, together with their functional requirements, need to be provided by the developers in a conventional manner through the Cloud layer. For example, the components of a real-time video monitoring and analysis application can be provided through the Cloud layer. Accordingly, the novel concepts, methodologies and technologies introduced in this work are transparent for the IoT applications.   

The SmartFog architecture evolves around the Fog layer, which can potentially enable autonomous management and provisioning of the computational and storage resources to the IoT applications deployed through the Cloud layer. This layer utilizes Spectral clustering techniques to perform efficient \textit{self-clustering} of the Fog devices, therefore emulating the function of the human brain. The clustering process is autonomously executed by a large set of specialized Fog devices called \textit{communication gateways} (depicted in yellow on Figure \ref{fig:system}), that also act as main points for communication and decision making between the Cloud and the Fog, thus emulating the function of the brain's \textit{corpus callosum}. For the selection of communication gateways or neurons, the Cloud layer utilizes \textit{multi-criteria sorting and decision making algorithms} by considering three distinctive criteria: betweenness centrality, computational performance, and communication latency to the Cloud. The utilization of the concept of betweenness centrality, as an objective in the multi-criteria sorting algorithm, enables identification of the topologically most central Fog devices with sufficient amount of resources, capable of performing the self-clustering of the Fog devices.  

Afterwards, the selected communication gateways collaboratively cluster the remaining Fog devices, so-called neurons, into specific \textit{functional areas} (depicted with colored dashed lines), thus organizing the Fog similar to a human brain. From a strictly technical point of view, the functional areas are clusters, or more concretely connected graphs, of logically grouped Fog neurons, capable of optimally supporting specific types of IoT applications. The clustering process of the functional areas is conducted by considering multiple performance related criteria. Therefore, each functional area is optimized for a specific application type, including computation, memory and network-intensive. The clustering process is continuously conducted in an adaptive manner that reacts to changes in the environment, such as unforeseen changes in the architecture (addition or removal of Fog devices). To enable fast clustering in distributed environments, we utilize the Spectral clustering technique, described in Section \ref{sec:background}.

\subsection{Communication gateways selection}
The process of Communication gateways selection is of utmost importance for reducing the decision making latency and increasing the resources utilization. Essentially, the Communication gateways act as a high-bandwidth aggregation points between the edge and the Cloud. Furthermore, they perform essential operations for clustering of the Fog devices in functional areas and mapping of the IoT applications to the physical resources. 
The selection of the Communication gateways is performed by the Cloud layer, which sorts the available Fog devices based on dominance in multi-dimensional search space. The approach that we take to perform the non-domination sorting is based on the Fast Elitist Non-dominated Sorting Genetic Algorithm (NSGA-II) \cite{deb2000fast}. In this approach, every Fog device from the full set of available devices, is checked with a partially filled set $F'$. To begin with, the first Fog device from the list of devices is initially kept in the set $F'$. Afterwards, each Fog device $f$ is compared with all members of the set $F'$ one by one. If the Fog device $f$ dominates any other fog device $q$ in the set $F'$, then the device $q$ is removed from the set. On the other hand, if device $f$ is dominated by all member of $F'$, then $f$ is ignored. Moreover, if device $f$ is not dominated by any device in $F'$, then it is included in $F'$. When all Fog devices known to the cloud are checked, the remaining members of $F'$ constitute the non-dominated set. 

The non-domination sorting algorithm is performed by considering the following conflicting criteria: (i) betweenness centrality, (ii) computational performance, and (iii) communication latency to the Cloud. The main purpose of the non-domination soring is to identify the Fog devices with highest number of available resources and the most optimal position in the Fog system. The betweenness centrality, as one of the objectives in the non-domination sorting algorithm, provides information on how well the each Fog device is connected in relation to the other Fog devices, the Cloud layer above and the IoT layer below. Respectively, the computational performance objective provides information on the computational capacity of the Fog devices. For the purposes of the SmartFog system, the computational performance is represented through the number of instructions that the processor of the Fog device can execute in one second - MIPS. The communication latency objective is represented through the virtual link latency between a given Fog device and any other device. In our implementation, historical data on the previous data transfers is being preserved to calculate the average latency between any two devices in the Fog environment.

The process of non-domination sorting algorithm results with a set of multiple optimal solutions, visualized in the form of Pareto front. As all of the Fog devices represented in the Pareto front are optimal, we utilize automated decision making strategy, which considers a-priory input information to select only the required number of Fog devices to act as Communication gateways. The decision making algorithm considers the number $N_f$ and the type $F$ of the functional areas to give priority which Fog devices to be selected. For example, if a given functional area should be optimized for processing, then the decision making algorithm will select the Fog device with the highest computational performance from the set $F'$ of Pareto optimal Fog devices. The selection process of the Communication gateways is presented in Algorithm \ref{alg:sort}.       
\begin{algorithm}
\label{alg1}
\caption{Communication gateways selection}\label{alg:sort}
\begin{algorithmic}[1]
\INPUT
\Statex $N$ \Comment{Number of Fog devices}
\Statex $L = (L_1, L_2, ... , L_N)$ \Comment{List of Fog devices}
\Statex $C = (C_1, C_2, ... , C_N)$ \Comment{CPUs per Fog device}
\Statex $M = (M_1, M_2, ... , M_N)$ \Comment{Memory per Fog device}
\Statex $T$ \Comment{Topology of the Fog layer}
\Statex $N_f$ \Comment{Number of functional areas}
\Statex $F = (F_1, F_2, ..., F_{N_f} )$ \Comment{Type of the functional area}
\While{$i < N$}
\State $B_i \gets evaluate\_betweenness\_centrality(L_i, T)$
\State $E_i \gets evaluate\_Fog\_device(C_i, M_i,B_i)$
\State $i \gets i+1$
\EndWhile\label{mainloop}
\State $S \gets non\_domination\_sorting(E)$
\State $D \gets automated\_decision\_making (S, N_f,F)$
\end{algorithmic}
\end{algorithm}

\subsection{Functional areas clustering}
The concept of functional areas, introduced by the SmartFog architecture, evolves around the notion of Fog devices' grouping into distinctive clusters optimized for special types of IoT applications, such as compute or memory intensive. Essentially, the clustering process enables grouping of logically similar Fog devices by considering various resource related criteria. 

The clustering process and the formation of the functional areas is performed in distributed manner among the communication gateways. Every communication gateway performs Spectral clustering by considering the available computing and memory resources of the available Fog devices. A-priory input from the Cloud layer is used to steer the clustering process in the preferred direction, therefore creating clusters of Fog devices with similar amount of available processing and memory resources. For example, if a communication gateway is intended to create compute optimized functional area, the Spectral clustering will identify all similar Fog devices, which have sufficient amount of processing resources. The process of functional areas formation is presented in Algorithm \ref{alg:cluster}. It is essential to be noted, that every communication gateway creates each own cluster, therefore a given Fog device can be member of multiple functional areas, provided it has sufficient resources.  

\begin{algorithm}
\caption{Functional areas clustering}\label{alg:cluster}
\begin{algorithmic}[1]
\INPUT
\Statex $N$ \Comment{Number of Fog devices}
\Statex $k$ \Comment{Number of clusters}
\Statex $C = (C_1, C_2, ... , C_N)$ \Comment{CPUs per Fog device}
\Statex $M = (M_1, M_2, ... , M_N)$ \Comment{Memory per Fog device}
\Statex $G$ \Comment{Value of the Gaussian filter}
\State $L \gets create\_dense\_matrix(C, M)$
\State $Y \gets find\_eigen\_vectors(L, k)$
\State $K_m \gets apply\_k\_means(Y, k)$
\State $F_k \gets create\_functional_clusters (K_m)$

\end{algorithmic}
\end{algorithm}

%% file: sections/Implementation.tex
In this section, we discuss the essential implementation details of the proposed Fog architecture with respect to the multi-criteria non-domination sorting and spectral clustering algorithms.

\subsection{Simulation environment} \label{sec:over}
We simulated an unstructured Fog overlay network system on top of the iFogSim \cite{gupta2017ifogsim}, which is an efficient toolkit for modeling and simulating resource management techniques in IoT and Fog computing environments. To accommodate the novel concepts introduced by SmartFog we have extended iFogSim to support the automated gateways selection and devices' clustering.

Within the simulation environment we assume that every Fog node in the overlay network have a specific processing capacity and system architecture, such as ARM or x86. Furthermore, every node has limited operating memory and storage disk capacity with a fixed number of stored data items of varying size. In addition a specific scheme has been defined to allow every Fog node to leave and join the network after a random time interval, which is an important scenario supported by the SmartFog architecture. 

\subsection{Multi-criteria decision making and spectral clustering}
\label{sec:optimization}
We perform the gateway selection based on multiple conflicting objectives, which have been modeled as a part of the information collected from the initial Fog overlay simulation. We utilize modified  NSGA-II multi-objective optimization algorithm to perform the non-domination sorting of the Fog devices. In addition to this, we also utilize low-latency automated decision making algorithm. To instantiate NSGA-II as a main component of the communication gateway selection module, we have extended the jMetal \cite{durillo2011jmetal} object-oriented Java framework for multi-objective optimization problems. We have implemented particular modifications in the jMetal framework to deal with the specific characteristics of the proposed architecture. More concretely, we have modified the non-domination sorting algorithm, as the standard operators of the jMetal framework do not support independent multi-criteria sorting. Furthermore, jMetal was extended to support automated decision-making by utilizing the algorithm proposed in \cite{dm}. This algorithm performs low-latency decision making by dividing the Pareto front in multiple regions based on the a-priory knowledge that gives priority to specific criteria. Therefore, this algorithm is suitable for the SmartFog architecture. Moreover, we modified the jMetal framework and extended its API to support integration with the iFogSim simulation environment. 

Lastly, the Spectral clustering was implemented by extending the WEKA data mining framework \cite{hall2009weka}. The implementation of the spectral clustering was based on the algorithm proposed in \cite{ng2002spectral}. Additionally, the Spectral clustering implementation was integrated within the jMetal framework, thus fully supporting the novel concepts introduced in the SmartFog. 

%% file: sections/Evaluation.tex
In this section we present broad experimental evaluation of the proposed concepts for Fog architecture management. We conducted the evaluation based on a set of indicators, specifically selected for the analysis of the non-domination sorting and clustering operations, covering: (i) the communication gateway selection, and (ii) the functional area grouping. Furthermore, we conducted extensive simulation-based evaluation to investigate the influence of the introduced concepts in the overall efficiency of the Fog architectures.  

\subsection{Communication gateway non-domination sorting and decision making}
\label{cg}
The essential feature of the SmartFog architecture is the introduction of the communication gateways, which act both as a aggregation points between the Edge and the Cloud, and are actively involved in the creation of the functional areas. For those reasons, it is important to evaluate the behavior, efficiency and scalability of the implemented non-domination sorting and decision making algorithms. For the evaluation purposes we utilize network overlays with sizes varying from 20 to 40 vertices, i.e. Fog devices. In the simulation test-bed we assumed that the Fog devices have limited computation resources, ranging from 800 to 1200 MIPS. In relation to the memory and storage resources, we assume that each device can have between 1 and 4 GB of operating memory and small amount of solid-state based storage. All experiments were repeated 100 times and the median value, together with the standard deviation, are presented in the Figures below.   

\begin{figure}[h]
    \centering
    \includegraphics[width=0.4\textwidth]{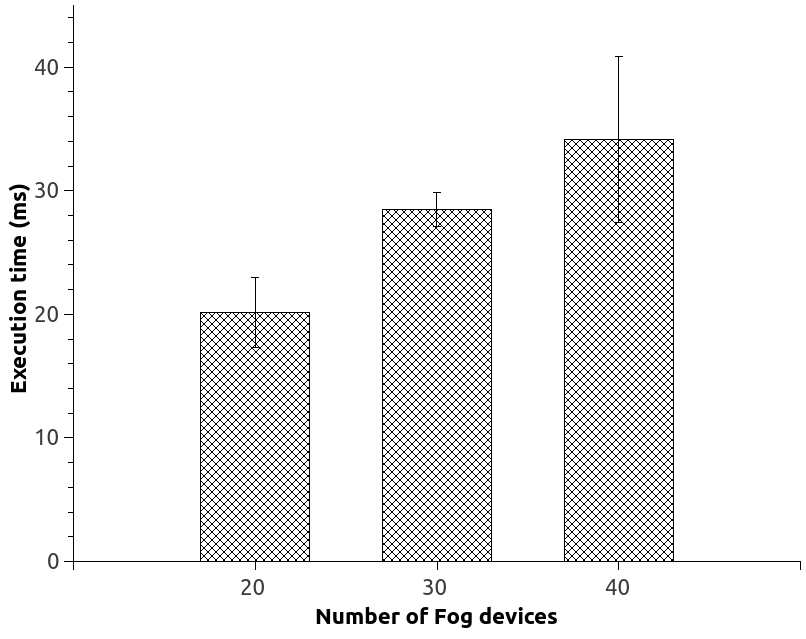}
    \caption{Computation scalability of the betweenness centrality algorithm}
    \label{fig:Between}
\end{figure}

Initially, the evaluation activities were focused towards validation of the concepts, introduced from the areas of graph theory and multi-criteria decision making. Therefore, we first explore the connectivity characteristics, expressed as betweenness centrality, of every Fog device, which are utilized in the multi-criteria non-domination sorting. The exploration of the scalability potential of the betweenness centrality algorithm is essential, as it should be executed each time there are changes in the highly volatile IoT/Fog network overlay. Moreover, this process requires calculation of all shortest path routes in a given topology, therefore increasing the complexity of the algorithm. The evaluation of the the betweenness centrality algorithm is presented on Figure \ref{fig:Between}.   

As it can be observed from the simulation results, the execution time of the betweenness centrality calculation rises linearly with the size of the network overlay. Nevertheless, even for 40 Fog devices, which can potentially serve thousands of Edge and IoT devices, the execution time is below 40 $ms$. 

\begin{figure}[h]
    \centering
    \includegraphics[width=0.4\textwidth]{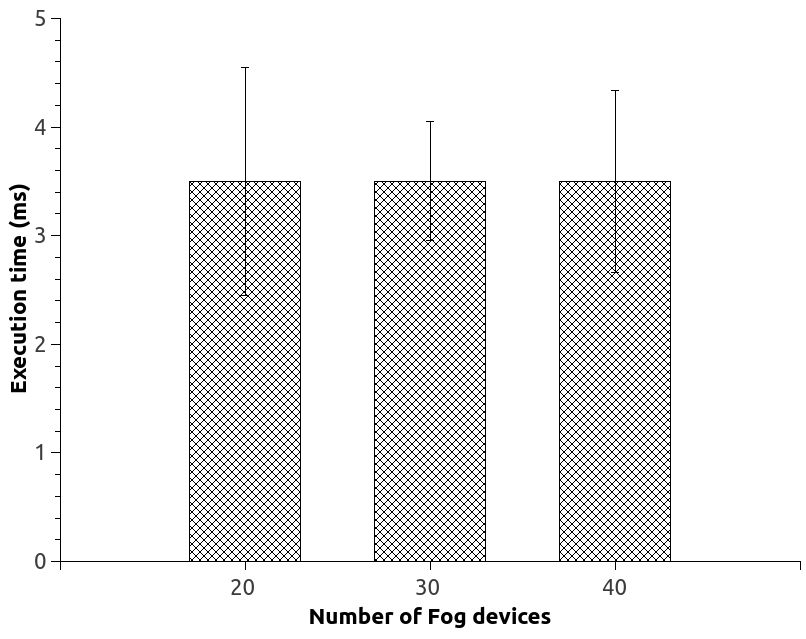}
    \caption{Computation scalability of the Non-domination sorting algorithm}
    \label{fig:Nondom}
\end{figure}

Furthermore, on Figure \ref{fig:Nondom}, we provide evaluation data of the non-domination sorting algorithm. The experimental scenario is identical with the one described above. The execution time presented in the Figure includes the time required for the decision making to be performed, while omitting the betweenness centrality calculation. We can observe that for our testbed, the non-domination sorting was very efficient and induced latencies below 5 $ms$.   

\subsection{Functional areas clustering}
\begin{figure}[h]
    \centering
    \includegraphics[width=0.4\textwidth]{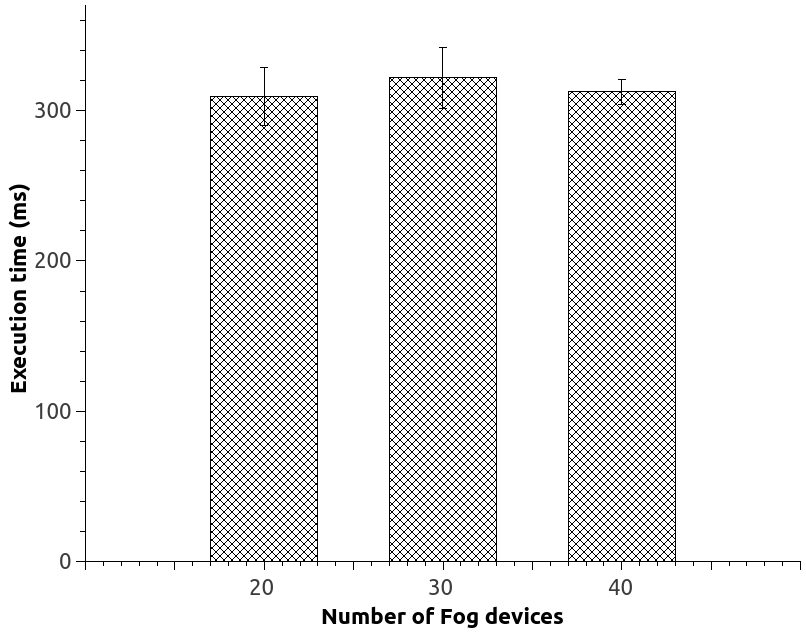}
    \caption{Computation scalability of the Spectral clustering algorithm}
    \label{fig:Spectral}
\end{figure}

\begin{figure}[h]
    \centering
    \includegraphics[width=0.4\textwidth]{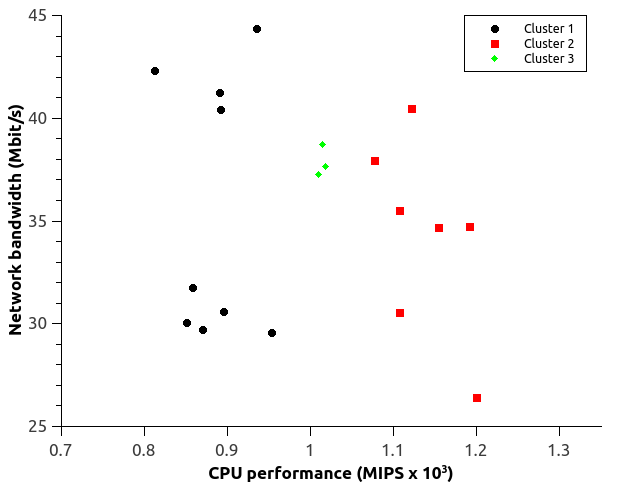}
    \caption{Sample results from the functional areas clustering}
    \label{fig:clus}
\end{figure}

The other concept, exploited by the SmartFog architecture, is the Spectral clustering algorithm, which is being utilized by the communication gateways to create specialized functional areas. The current implementation of the SmartFog architecture supports the creation of computational and memory optimized functional areas. For the purpose of Spectral clustering we again consider varying sizes of the network, which ranges from 20 to 40 Fog devices. The evaluation results, presented on Figure \ref{fig:Spectral}, clearly show that the algorithm scales very well, with latencies in the range of 300 $ms$ and very low standard deviation values. Moreover, on Figure \ref{fig:clus} sample clustering result for three functional areas is presented. 

\subsection{Simulation experiment}
In order to explore the influence of the introduced concepts on the overall efficiency of the Fog architectures we have conducted extensive simulation-based analysis. We have simulated a Sense-Process-Actuate Model (SPAM) with Edge-ward placement strategy. In SPAM, the sensors continuously emit data in the form of tuples. Tuple is basic unit of communication between different entities in iFogSim, which is specified by the following parameters: data source, data destination and amount of required computation and network resources. These tuples are transmitted to the Fog devices as data streams. The IoT applications modules, executed on the Fog devices, process this data and transmit it back to the actuators for performing a specific action. For the Edge-ward placement strategy, specific application's modules are deployed at edge of the network, while the remaining modules are placed to act as communication gateways to the Cloud. Therefore part of the data is being processed by Fog devices placed at edge of network and the remaining data is transmitted to the Cloud for further processing.  

\begin{figure}[h]
    \centering
    \includegraphics[width=0.4\textwidth]{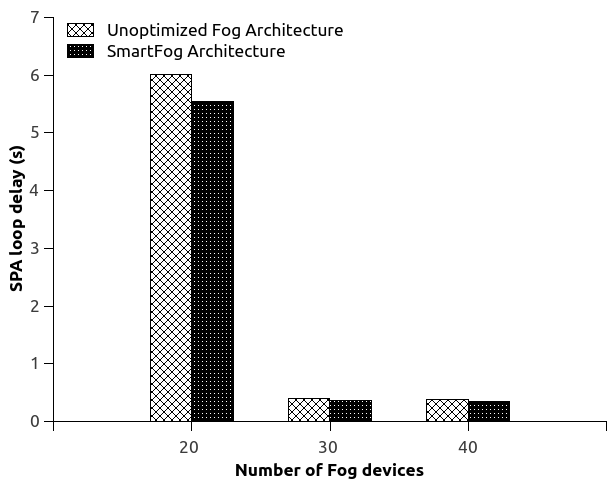}
    \caption{SPA loop latency in Fog environment}
    \label{SPA}
\end{figure}

\begin{figure}[h]
    \centering
    \includegraphics[width=0.4\textwidth]{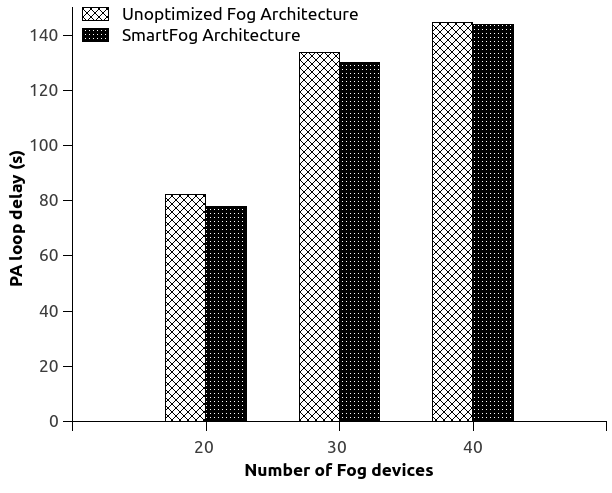}
    \caption{PC loop latency in Fog environment}
    \label{PC}
\end{figure}

In our simulation model, the sensors and actuators are attached to the edge devices (mobiles). These mobiles are connected with the Fog devices through edge-gateways. Whereas on the other end, the Fog devices are connected with the Cloud through the communication gateways. We have used edge-ward placement strategy in our simulation. Therefore we have defined two process-control loops to calculate the end-to-end latency or IoT application loop delay. The first is a sense-process-actuate (SPA) loop in which the Fog devices receive the data from the sensors, and then process it and transmit it back to the actuators for further action. The second one is the process-control (PC) loop in which, through communication gateways, the Cloud receives data from the Fog devices, and then process it and transmits it back to the source Fog devices.  Therefore, we conducted a series of experiments to evaluate the performance of the SmartFog against unoptimized-Fog in relation to the SPA and PC loop delays. These series consisted of three experimental scenarios with three overlay topologies with sizes of 20, 30, and 40 Fog devices respectively. The Fog devices in the overlays are connected randomly with one another creating a mesh kind of topology. Furthermore, the Fog devices are connected with the sensors and actuators, which are continuously emitting tuples. The Fog devices are also attached to Cloud through the communication gateways. For calculating the SPA loop delay, we have used tuples with a requirement of 1000-8000 MIPS and 100 bytes network width. Moreover, for the PC loop delay, the tuples have minimal requirement of 40000 MIPS and 100 bytes. In our experiments the tuples are sent randomly from the sensors to the Fog devices, while SmartFog manages the data transfers from the Fog to the Cloud. The simulation results from the loop delay evaluation are presented on Figures \ref{SPA} and \ref{PC}.


The evaluation results, both for the SPA and the PC loop latency, clearly show that SmartFog can reduce the decision making latency by up to 8\%, which is significant improvement in volatile environments such as the Fog. In the cases when the data needs to be processed in the Cloud, more concretely within the PC loop, we can conclude that for smaller overlays SmartFog can reduce the communication latency, while for larges overlays the latency remains the same. Furthermore, in the case of the SPA loop, SmartFog can significantly reduce the latency, both for small and large overlays.  

\begin{figure}[h]
    \centering
    \includegraphics[width=0.4\textwidth]{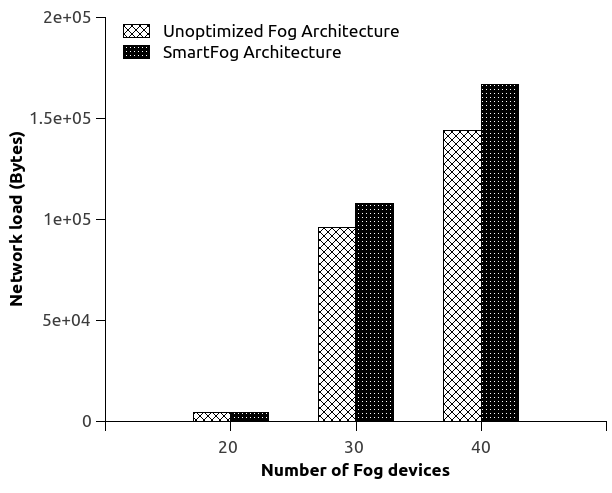}
    \caption{Network load in Fog environment}
    \label{Nload}
\end{figure}

Moreover, we have further focused our experimental evaluation on the network load induced by the IoT applications. For this reason, we have evaluated the total network traffic over the overlays for the SmartFog architecture and unoptimized Fog. The results, presented on Figure \ref{Nload}, clearly show that for all network sizes the communication load can be reduced by up to 13\%.    

%% file: sections/Conclusion.tex
In this paper we introduce a promising novel nature-inspired Fog architecture, named SmartFog, capable of providing low decision making latency and adaptive architecture management. By utilizing novel algorithms and techniques from the fields of multi-criteria decision making, graph theory and machine learning  we model the Fog as a distributed intelligent processing system, therefore emulating the function of the human brain. One of the key strengths of our approach are the ability (i) to identify the most optimal aggregation points between the Cloud and the Edge, and (ii) to adaptively group the Fog devices in optimized clusters based on similarity.
We have implemented, integrated and evaluated the introduced concepts as essentials elements of the SmartFog environment. Based on the evaluation results, it can be concluded that the proposed nature-inspired Fog architecture can provide up to 8\% latency reduction, and 13\% reduction in network load.  


Regarding the future research activities, we plan to extended the current environment to support proximity aware data management and adaptive resource provisioning.

%% file: main.bbl
\begin{thebibliography}{10}

\bibitem{mainEdge}
Mahadev Satyanarayanan.
\newblock The emergence of edge computing.
\newblock {\em Computer}, 50(1):30--39, 2017.

\bibitem{2016fog}
Amir~Vahid Dastjerdi, Harshit Gupta, Rodrigo~N Calheiros, Soumya~K Ghosh, and
  Rajkumar Buyya.
\newblock Fog computing: Principles, architectures, and applications.
\newblock {\em arXiv preprint arXiv:1601.02752}, 2016.

\bibitem{varshney2017demystifying}
Prateeksha Varshney and Yogesh Simmhan.
\newblock Demystifying fog computing: Characterizing architectures,
  applications and abstractions.
\newblock {\em arXiv preprint arXiv:1702.06331}, 2017.

\bibitem{orchFog}
Zhenyu Wen, Renyu Yang, Peter Garraghan, Tao Lin, Jie Xu, and Michael Rovatsos.
\newblock Fog orchestration for internet of things services.
\newblock {\em IEEE Internet Computing}, 21(2):16--24, 2017.

\bibitem{vaquero2014finding}
Luis~M Vaquero and Luis Rodero-Merino.
\newblock Finding your way in the fog: Towards a comprehensive definition of
  fog computing.
\newblock {\em ACM SIGCOMM Computer Communication Review}, 44(5):27--32, 2014.

\bibitem{fogarch}
Flavio Bonomi, Rodolfo Milito, Preethi Natarajan, and Jiang Zhu.
\newblock Fog computing: A platform for internet of things and analytics.
\newblock In {\em Big Data and Internet of Things: A Roadmap for Smart
  Environments}, pages 169--186. Springer, 2014.

\bibitem{kimovski2015parallel}
Dragi Kimovski, Julio Ortega, Andr{\'e}s Ortiz, and Ra{\'u}l Ba{\~n}os.
\newblock Parallel alternatives for evolutionary multi-objective optimization
  in unsupervised feature selection.
\newblock {\em Expert Systems with Applications}, 42(9):4239--4252, 2015.

\bibitem{centrality}
Ulrik Brandes, Stephen~P Borgatti, and Linton~C Freeman.
\newblock Maintaining the duality of closeness and betweenness centrality.
\newblock {\em Social Networks}, 44:153--159, 2016.

\bibitem{spectral2007tutorial}
Ulrike Von~Luxburg.
\newblock A tutorial on spectral clustering.
\newblock {\em Statistics and computing}, 17(4):395--416, 2007.

\bibitem{shoker2016lightkone}
Ali Shoker, Joao Leitao, Peter Van~Roy, and Christopher Meiklejohn.
\newblock Lightkone: Towards general purpose computations on the edge.
\newblock White Paper published on http://www.lightkone.eu, 2016.

\bibitem{plebani2017information}
Pierluigi Plebani, David Garcia-Perez, Maya Anderson, David Bermbach, Cinzia
  Cappiello, Ronen~I Kat, Frank Pallas, Barbara Pernici, Stefan Tai, and Monica
  Vitali.
\newblock Information logistics and fog computing: The ditas approach.
\newblock In {\em CAiSE 2017 Forum}, 2017.

\bibitem{osmotic}
Massimo Villari, Maria Fazio, Schahram Dustdar, Omer Rana, and Rajiv Ranjan.
\newblock Osmotic computing: A new paradigm for edge/cloud integration.
\newblock {\em IEEE Cloud Computing}, 3(6):76--83, 2016.

\bibitem{OrchFogData}
Mathias~Santos de~Brito, Saiful Hoque, Thomas Magedanz, Ronald Steinke,
  Alexander Willner, Daniel Nehls, Oliver Keils, and Florian Schreiner.
\newblock A service orchestration architecture for fog-enabled infrastructures.
\newblock In {\em Fog and Mobile Edge Computing (FMEC), 2017 Second
  International Conference on}, pages 127--132. IEEE, 2017.

\bibitem{aazam2015dynamic}
Mohammad Aazam and Eui-Nam Huh.
\newblock Dynamic resource provisioning through fog micro datacenter.
\newblock In {\em Pervasive Computing and Communication Workshops (PerCom
  Workshops), 2015 IEEE International Conference on}, pages 105--110. IEEE,
  2015.

\bibitem{OrchIoT}
Olena Skarlat, Stefan Schulte, Michael Borkowski, and Philipp Leitner.
\newblock Resource provisioning for iot services in the fog.
\newblock In {\em Service-Oriented Computing and Applications (SOCA), 2016 IEEE
  9th International Conference on}, pages 32--39. IEEE, 2016.

\bibitem{branke2008multiobjective}
J{\"u}rgen Branke, Kalyanmoy Deb, and Kaisa Miettinen.
\newblock {\em Multiobjective optimization: Interactive and evolutionary
  approaches}, volume 5252.
\newblock Springer Science \& Business Media, 2008.

\bibitem{kimovski2016multi}
Dragi Kimovski, Nishant Saurabh, Sandi Gec, Vlado Stankovski, and Radu Prodan.
\newblock Multi-objective optimization framework for vmi distribution in
  federated cloud repositories.
\newblock In {\em European Conference on Parallel Processing}, pages 236--247.
  Springer, 2016.

\bibitem{barthelemy2004betweenness}
Marc Barthelemy.
\newblock Betweenness centrality in large complex networks.
\newblock {\em The European Physical Journal B-Condensed Matter and Complex
  Systems}, 38(2):163--168, 2004.

\bibitem{cluster}
Rui Xu and Donald Wunsch.
\newblock Survey of clustering algorithms.
\newblock {\em IEEE Transactions on neural networks}, 16(3):645--678, 2005.

\bibitem{edelman1988eigenvalues}
Alan Edelman.
\newblock Eigenvalues and condition numbers of random matrices.
\newblock {\em SIAM Journal on Matrix Analysis and Applications},
  9(4):543--560, 1988.

\bibitem{kmeans}
Tapas Kanungo, David~M Mount, Nathan~S Netanyahu, Christine~D Piatko, Ruth
  Silverman, and Angela~Y Wu.
\newblock An efficient k-means clustering algorithm: Analysis and
  implementation.
\newblock {\em IEEE transactions on pattern analysis and machine intelligence},
  24(7):881--892, 2002.

\bibitem{deb2000fast}
Kalyanmoy Deb, Samir Agrawal, Amrit Pratap, and Tanaka Meyarivan.
\newblock A fast elitist non-dominated sorting genetic algorithm for
  multi-objective optimization: Nsga-ii.
\newblock In {\em International Conference on Parallel Problem Solving From
  Nature}, pages 849--858. Springer, 2000.

\bibitem{gupta2017ifogsim}
Harshit Gupta, Amir Vahid~Dastjerdi, Soumya~K Ghosh, and Rajkumar Buyya.
\newblock ifogsim: A toolkit for modeling and simulation of resource management
  techniques in the internet of things, edge and fog computing environments.
\newblock {\em Software: Practice and Experience}, 47(9):1275--1296, 2017.

\bibitem{durillo2011jmetal}
Juan~J Durillo and Antonio~J Nebro.
\newblock jmetal: A java framework for multi-objective optimization.
\newblock {\em Advances in Engineering Software}, 42(10):760--771, 2011.

\bibitem{dm}
Roland~Matha Dragi~Kimovski, Sasko~Ristov and Radu Prodan.
\newblock Multi-objective service oriented network provisioning in ultra-scale
  systems.
\newblock In {\em EUROPAR Workshops 2017}. Springer, 2017.

\bibitem{hall2009weka}
Mark Hall, Eibe Frank, Geoffrey Holmes, Bernhard Pfahringer, Peter Reutemann,
  and Ian~H Witten.
\newblock The weka data mining software: an update.
\newblock {\em ACM SIGKDD explorations newsletter}, 11(1):10--18, 2009.

\bibitem{ng2002spectral}
Andrew~Y Ng, Michael~I Jordan, and Yair Weiss.
\newblock On spectral clustering: Analysis and an algorithm.
\newblock In {\em Advances in neural information processing systems}, pages
  849--856, 2002.

\end{thebibliography}
